\documentclass{aastex}
\usepackage{natbib}
\usepackage{hyperref}
\usepackage{multirow}
\usepackage{lscape}

\hypersetup{colorlinks=true, breaklinks=true, citecolor=blue, linkcolor=blue, menucolor=blue, urlcolor=blue}

\newcommand{\so}{PKS~2004-447}
\newcommand{\sy}{NLSy1s}
\newcommand{\nls}{NLSy1}
\newcommand{\ser}{S\'ersic}
\newcommand{\dev}{de~Vaucouleurs}

\begin{document}

\title{Discovery of a pseudobulge galaxy launching powerful relativistic jets} 

\author{Jari K. Kotilainen\altaffilmark{1,2}}
\affil{Finnish Centre for Astronomy with ESO (FINCA), University of Turku, V\"ais\"al\"antie 20,FI-21500  Piikki\"o, Finland}

\author{Jonathan Le{\'o}n Tavares\altaffilmark{3}}
\affil{Sterrenkundig Observatorium, Universiteit Gent, Krijgslaan 281-S9, B-9000 Gent, Belgium}

\author{Alejandro Olgu{\'{\i}}n-Iglesias\altaffilmark{4}}
\affil{Instituto Nacional de Astrof\'{\i}sica \'Optica y Electr\'onica (INAOE), Apartado Postal 51 y 216, 72000 Puebla, Mexico}

\author{Maarten Baes}
\affil{Sterrenkundig Observatorium, Universiteit Gent, Krijgslaan 281-S9, B-9000 Gent, Belgium}

\author{Christopher An{\'o}rve}
\affil{Facultad de Ciencias de la Tierra y del Espacio de la Universidad Aut\'onoma  de Sinaloa, Blvd. de la Americas y Av. Universitarios S/N, Ciudad Universitaria, C.P. 80010, Culiac\'an Sinaloa, M\'exico}

\author{Vahram Chavushyan}
\affil{Instituto Nacional de Astrof\'{\i}sica \'Optica y Electr\'onica (INAOE), Apartado Postal 51 y 216, 72000 Puebla, Mexico}

%% Use the \and command so offset the last author.
\and

\author{Luis Carrasco}
\affil{Instituto Nacional de Astrof\'{\i}sica \'Optica y Electr\'onica (INAOE), Apartado Postal 51 y 216, 72000 Puebla, Mexico}

\altaffiltext{1}{jarkot@utu.fi}
\altaffiltext{2}{also at: Tuorla Observatory, Department of Physics and Astronomy, University of Turku, V\"ais\"al\"antie 20,FI-21500  Piikki\"o, Finland}
\altaffiltext{3}{also at: Instituto Nacional de Astrof\'{\i}sica \'Optica y Electr\'onica (INAOE), Apartado Postal 51 y 216, 72000 Puebla, Mexico}
\altaffiltext{4}{also at: Finnish Centre for Astronomy with ESO (FINCA), University of Turku, V\"ais\"al\"antie 20,FI-21500  Piikki\"o, Finland}

\begin{abstract}
Supermassive black holes launching plasma jets at close to speed of light, producing gamma-rays, have ubiquitously been found to be hosted by massive elliptical galaxies. Since elliptical galaxies are generally believed to be built through galaxy mergers, active galactic nuclei (AGN) launching relativistic jets are associated to the latest stages of galaxy evolution. We have discovered a pseudo-bulge morphology in the host galaxy of the gamma-ray AGN PKS 2004-447.   This is the first gamma--ray emitter radio loud AGN found to be launched from a system where both black hole and host galaxy have been actively growing via secular processes. This is evidence for an alternative black hole-galaxy co-evolutionary path to develop powerful relativistic jets that is not merger-driven.

\end{abstract}

\section{Introduction}

Active galactic nuclei (AGN) launching relativistically moving plasma jets are classified as radio-loud. If its jet is pointing towards the Earth, the radio-loud AGN is called a blazar.  Relativistic jets in blazars are capable of producing bright and variable non-thermal emission across the entire electromagnetic spectrum \citep{giommi12}.  The large majority of high-Galactic latitude sources detected by the Large Area Telescope (LAT) onboard the Fermi Gamma-ray Space Telescope have been associated to blazars \citep{acker15}. Based on their optical spectra, blazars are classified as either flat-spectrum radio-quasars (FSRQ) showing prominent emission lines, or BL Lac objects (BL Lacs) with featureless spectra. Yet both blazar types are ubiquitously hosted by massive red elliptical galaxies \citep{kotilainen98,falomo00}.  In particular, the distribution of host galaxy magnitudes of BL Lacs follows a narrow Gaussian distribution and BL Lac host galaxies can therefore be assumed to be standard candles when estimating their distances \citep{kotilainen11}. It is therefore well established that gamma-ray emission from AGN is associated to relativistic jets of blazars launched from the centres of massive elliptical galaxies widely believed to be formed by galaxy mergers. 

This scenario has been challenged by the detection of gamma-ray emission from Narrow-Line Seyfert 1 (NLSy1) galaxies \citep{abdo09,dammando15} that as a class are predominantly radio-quiet  and hosted in lower-mass spiral galaxies \citep{deo06,ohta07}.  The optical emission-line spectra of NLSy1s resemble those of Seyfert type 1 galaxies, with the exception of showing remarkably narrow (FWHM $< 1000$ km s$^{-1}$) Balmer emission lines. This, together with their cast signatures of large variability in X-rays and prominent Fe emission have contributed to the idea that the central engines of NLSy1s are powered by extreme accretion (rates close to the Eddington limit) onto a young and undermassive black hole that is still growing. Several studies have addressed the multi-wavelength emission of the six   gamma-ray NLSy1s so far detected by Fermi, converging to the idea that gamma-ray NLSy1 nuclei are blazar like, albeit less powerful \citep{foschini15}. Little is, however, still known about the host galaxies of the gamma-ray emitting NLSy1s \citep{leontavares14}.

In this work we present for the first time host galaxy images of the gamma-ray \nls\ \so,  with a redshift of $z=0.24$ \citep{drinkwater97} and showing core-jet parsec and kiloparsec jet scale structures, it is the radio--loudest \nls\ \citep[RL=1700--6300,][]{oshlack01} detected by Fermi/LAT \citep{orienti15,schulz16}. This manuscript is organised as follows.  In section 2 our observations are presented.  Section 3 describes the process and results from quantitatively characterising the host galaxy of \so. Our discussion is presented in section 4. Finally, in section 5 we summarise our findings.  Throughout the manuscript we adopt  cosmological parameters  of $\Omega_{m}=0.3$, $\Omega_{\Lambda} = 0.7$ and a Hubble constant of   $H_{0} = 70$ Mpc$^{-1}$ km s$^{-1}$.

\section{Observations and analysis}\label{sec:images}

Deep NIR ($J$ and $K_s$) images  of \so\ were obtained with ISAAC \citep[Infrared Spectrometer And Array Camera;][]{isaac}, mounted on UT3 of the VLT at the ESO in Paranal Chile. ISAAC is equipped with  a $1024 \times1024$ pixel Hawaii Rockwell array, with a pixel scale of 0.147 $"/$pixel, giving a field of view of $150" \times150"$. The observations were performed in service mode during the nights of 14 and 16  April 2013 with excellent atmospheric seeing conditions ($ FWHM\approx0.40"$ and $FWHM\approx0.45"$ for J-- and Ks--band, respectively). The total integration time was 600 s for J--band and 110 s for Ks--band. A jitter procedure was used to obtain a set of offset frames (with individual exposure times of 40 s and 10s for J-- and Ks--band, respectively) with respect to the initial position. Images were reduced with common procedures using IRAF. Each image was flat fielded and sky subtracted using a flat--frame derived from twilight images and a sky--frame derived by median filtering the individual frames in the stack (after scaling with the median of the number of counts in each image to account for sky intensity variations). The reduced images of each filter were aligned (using bright stars as reference points in the field) and combined to produce the final reduced co--added image. We have used  2MASS \citep{2mass} bright stars to calibrate the zeropoints of our J and Ks band images, resulting in an accuracy of $\sim$0.1 mag.

We analyse quantitatively the structure of our NIR images of \so\ by modeling their surface brightness distribution with  the 2D image decomposition code GALFIT \citep{peng11}.  The images were decomposed  into a combination of  a central point unresolved source (representing the AGN emission and modeled by a PSF) and galaxy components.  

\begin{figure}[t]
\includegraphics[width=\textwidth]{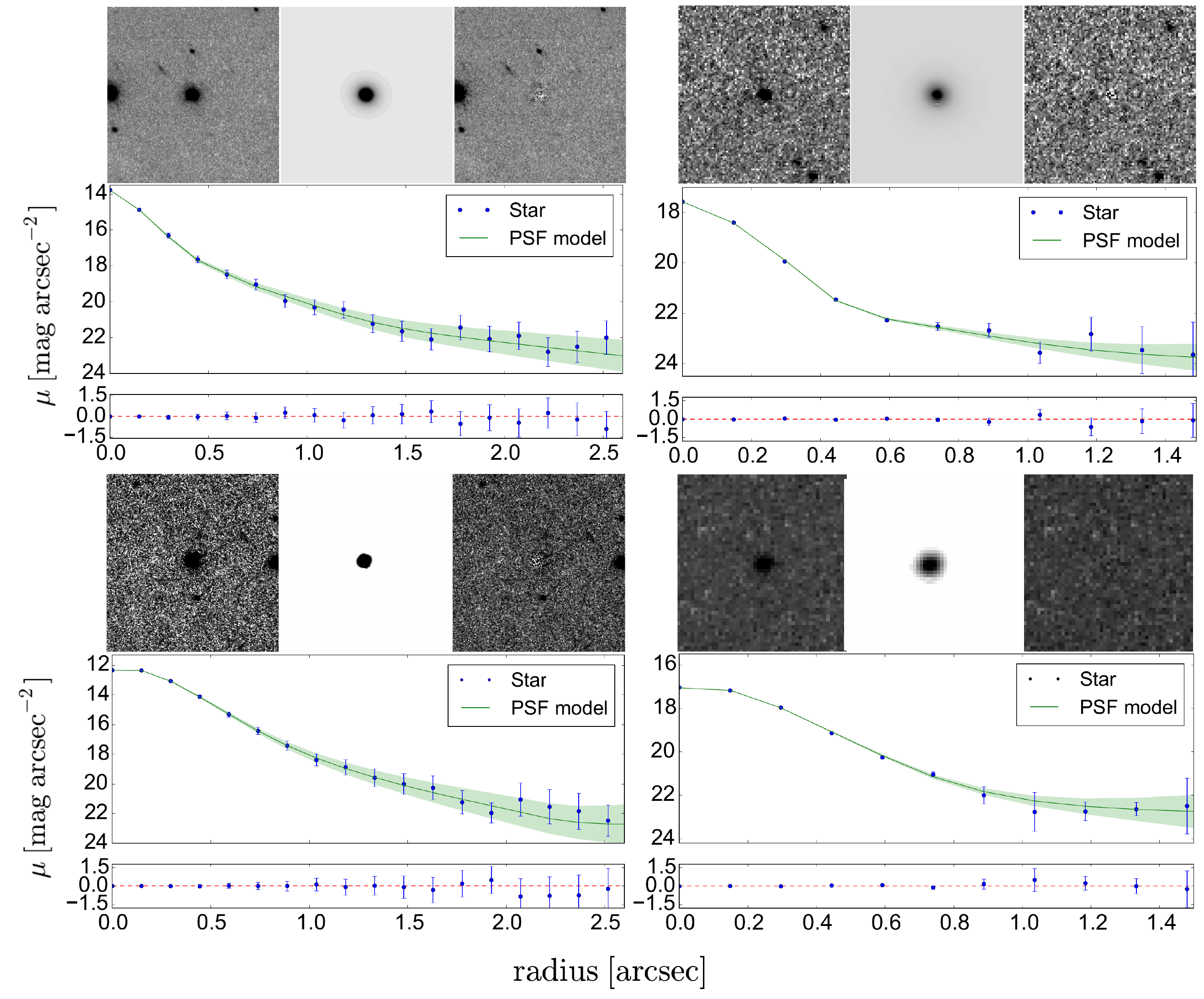}
\caption{Test to the PSF model. Each star is fitted with our
final PSF model in order to ensure its reliability. From left to right, the images in every panel are: the observed star, the PSF model and the residuals. The plots show the
azimuthally averaged surface brightness profiles of our PSF model and the fitted star. The
lower subpanels show the residuals of the fit. For every band we show a bright and a faint star. $J-$ and $Ks-$bands are shown in top and bottom row, respectively.}\label{psf_test}
\end{figure}

To model the PSF, we selected non--saturated field stars over a wide range of magnitude, without close companions (closer than $\sim 2.5"$) and preferentially close to the source. This resulted in a set of 10 stars with magnitudes $m_J=15.58^{+0.44}_{-0.39}$ and $m_K=15.04^{+0.47}_{-0.40}$, FWHM(J-band)$=0.40^{+0.012}_{-0.022}$ and FWHM(Ks-band)$=0.45^{+0.006}_{-0.010}$. We masked out unwanted objects by implementing the SExtractor segmentation image \citep{bertin96}, and fit all the selected stars simultaneously, using 2D modeling consisting of several Gaussian and exponential functions. Since our modeled PSF can fit adequately all the field stars (Figure \ref{psf_test}), we are confident that it is a good and stable representation of the true PSF of our images.\smallskip

Differences in the PSF due to spatial and temporal variations lead to uncertainties in the resulting parameters of our models. To account for this, we used different PSF models (derived from different combinations of stars), assuming that the resulting variations represent the uncertainties due to our PSF model imperfection. Additionally, our reported errors take into account the mismatch between our PSF and the stars used to model it, in this way, performing the fitting using empirical PSFs would result in models with components lying inside the error bars of the components modeled with our PSF.\smallskip

Uncertainties due to our PSF imperfection mainly affect the morphology of our models and to a lesser extent, their magnitudes. On the other hand, uncertainties due to the sky background can significantly affect the final magnitudes. To quantify this, we performed several sky fits in separate 100 px $\times$ 100 px regions of the field and ran our models with the mean ($\bar x$) and $\bar x\pm\sigma$ of the resulting sky values. The difference in magnitudes due to the uncertainties of the sky value is added to the errors of the final models. For a detailed description of the methodology to estimate uncertainties in the parameters of our modeling see \citet{leontavares14,olguin16}.\smallskip

%To model the PSF, we extract several  non-saturated stars from the field without close companions (closer than $\sim 2.5"$) and preferentially  located near the source. We mask out unwanted objects by implementing the Sextractor segmentation image \citep{bertin96}, and fit all the selected stars simultaneously, using 2D modeling consisting of Gaussian and exponential functions.  Since our modeled PSF can fit adequately any of the field stars in our images, we are confident that our PSF is good and stable  representation of the true PSF. 

%%%%%%%%%%%%%%%%%%%%%%%
%%%%%%%%%%%%%%%%%%%%%%%
%%%%%%%%%%%%%%%%%%%%%%%
\begin{table}
\center
\tiny
\caption{Model parameters from surface brightness decomposition. Columns are: (1)   \dev\ model (PSF + S\'ersic (n fixed to 4)), (2)  A three component model [PSF, S\'ersic + Exponential Disc model],  (3) a 4 component model [PSF,+S\'ersic+ exponential disc + bar], the addition of the bar component has been physically motivated by the identification of ansae (handles of the bar) in the residual images. N/H, B/H, D/H and bar/H refers to nucleus-, bulge-, disc-, and bar-to-host ratios, and $R_{s}$ is the disc scale length. Errors are shown enclosed in parentheses  below the associated parameter. Details on the surface brightness decomposition procedure and error estimation can be found in \cite{leontavares14}.}\label{tab:galfit}
\begin{tabular}{lccccccc}
\multirow{2}{*}{ }&\multicolumn{3}{c}{J band} &&\multicolumn{3}{c}{K band}	\\
\cline{2-4}
\cline{6-8}
&	1 	&	2	&	3 &&1 & 2 & 3 \\
& \multicolumn{7}{c}{Nucleus} \\
  \cline{2-8}
 \multirow{2}{*}{mag} & 17.69 & 17.87 & 17.84  && 15.45 &15.44  & 15.61 \\
& (0.40) &  (0.32) & (0.32) &&  (0.38) & (0.30) &  (0.25)\\ 
\multirow{2}{*}{N/H} &0.32 &0.28 &0.28  && 0.70  & 0.77 & 0.57 \\
&  (0.02)&  (0.03)& (0.02) && (0.04) &  (0.03) &   (0.03)\\
& \multicolumn{7}{c}{Bulge} \\
 \cline{2-8}
\multirow{2}{*}{ mag} & 16.45  &17.46  & 17.49   && 15.07  & 16.21   &15.90   \\
   &  (0.38) & (0.31) &   (0.29) &&  (0.35) &  (0.30) &  (0.28)   \\
\multirow{2}{*}{$R_{eff}$ [arcsec/kpc]} &0.82/3.12 & 0.18/0.70 & 0.19/0.72 && 0.60/2.28 &0.17/0.66 & 0.14 /0.53 \\
                  &(0.20/0.76) & (0.11/0.42) & (0.09/0.34) && (0.21/0.80) &(0.12/0.45) &(0.11/0.42) \\
\multirow{2}{*}{$n$}& 4.00 & 1.20 & 1.15 && 4.00 & 1.10  & 1.08 \\
                    &-- & (0.15)  & (0.10) && --&(0.41) &(0.35)\\
\multirow{2}{*}{$<\mu_{e}>$ [$mag/arcsec^2$]}& 17.26 &16.61 & 16.62 &&15.37 &15.17 &14.71    \\ 
                                       & (0.20) &(0.25) & (0.20) &&(0.19) &(0.21) &(0.19)\\ 
\multirow{2}{*}{B/H}& 1.00 & 0.41 &0.39 &&1.00 &0.38 &0.44   \\                
&(0.00) & (0.04) & (0.02) &&(0.00) &(0.03) &(0.03)   \\ 
\multirow{2}{*}{$\epsilon_{bulge}$}&0.57 & 0.52 & 0.54 && 0.63 & 0.46 & 0.48\\
                                   &(0.10) &(0.10) & (0.08) &&(0.11) & (0.14) & (0.12)\\
& \multicolumn{7}{c}{Disc}  \\
   \cline{2-8}
  \multirow{2}{*}{mag} &--& 17.06 & 17.35 &&-- & 15.67 & 16.00 \\		
  				&--&(0.30) &  (0.27) &&-- & (0.31) &(0.30) \\
\multirow{2}{*}{$R_{s}$ [arcsec/kpc]} & --& 0.81/3.07 & 0.96/3.65 && -- & 0.63/2.37 & 0.66/2.51 \\
								  & --& (0.12/0.45) & (0.08/0.32) && -- & (0.11/0.41) & (0.08/0.32) \\
\multirow{2}{*}{D/H} &--&0.59 & 0.44 &&-- & 0.62 & 0.40  \\
                     &--&  (0.04) &(0.03) && -- &(0.03) & (0.02)\\
\multirow{2}{*}{$\epsilon_{disc}$}&-- &0.58 & 0.71 &&--& 0.56& 0.75\\
                                  &-- &(0.12)  & (0.12)&&--&(0.11)&(0.16)\\
& \multicolumn{7}{c}{Bar}  \\
 \cline{2-8}
 \multirow{2}{*}{ mag}&-- & -- &18.39 && -- & -- & 16.96 \\
 							  &-- & -- & (0.31) &&-- & -- & (0.38)\\
 \multirow{2}{*}{ $R_{eff}$ [arcsec/kpc]} &-- & -- & 1.02/3.88 &&--&--& 1.07/4.06\\
 &-- & -- &(0.13/0.49) &&-- & -- & (0.13/0.49)        \\
\multirow{2}{*}{$n$}&-- & -- & 0.50 && -- & -- &0.88\\
&-- & -- &(0.20) && -- & -- &(0.22)\\
\multirow{2}{*}{Bar/H}&-- & -- & 0.17 && -- & -- & 0.16\\
&-- & -- & (0.02) && -- & -- & (0.02)\\
\multirow{2}{*}{$\epsilon_{bar}$}&-- & -- & 0.38 && -- & -- &0.27 \\
&-- & -- & (0.14)  && -- & -- &(0.16) \\
 \multirow{2}{*}{$\chi^2$}& 1.598 & 1.481 &1.403 && 1.618 & 1.483 &1.398\\
 &(0.011) &(0.025) & (0.031)&&(0.012) &(0.018) & (0.026)\\
\hline
\end{tabular}
\end{table}
%%%%%%%%%%%%%%%%%%%%%%%
%%%%%%%%%%%%%%%%%%%%%%%
%%%%%%%%%%%%%%%%%%%%%%%
%%%%%%%%%%%%%%%%%%%%%%%

\section{Host galaxy }

\begin{figure}[t]
\includegraphics[width=\textwidth]{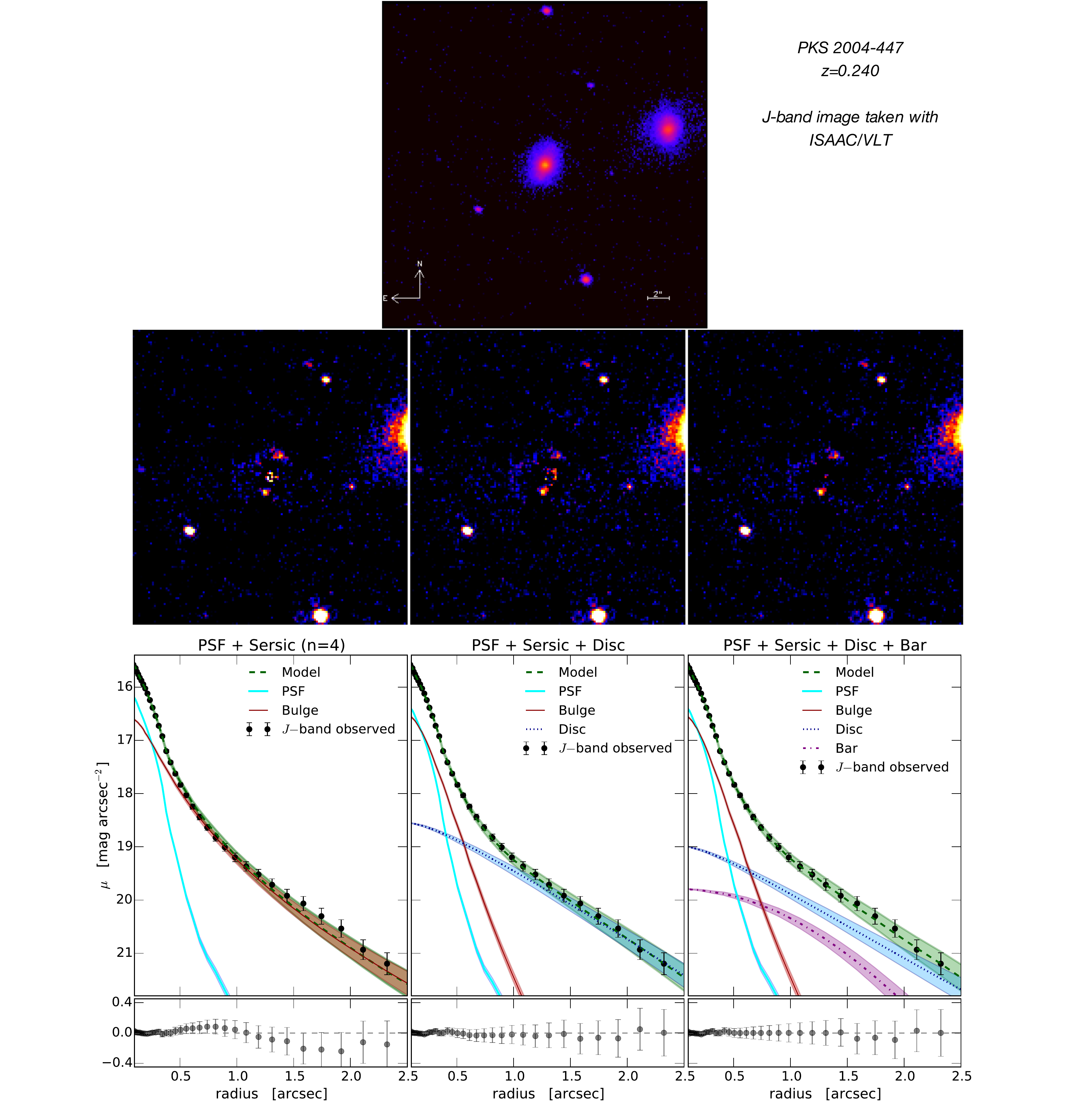}
\caption{Observed $J-$band images of \so. Residual image and azimuthally average surface brightness radial profiles for a de Vaucouleurs model (\textit{left panels}), a bulge with an exponential disc model (\textit{middle panels}) and a bulge with an exponential disc and a bar model (\textit{right panels}). Bottom panels show the radial distributions of the residuals. All three models comprise a nuclear unresolved source (PSF) to account for the AGN contribition. Note the presence of the two aligned spots in the residual images. This residual structures has been identified as the ansae of the bar.}\label{fig:j}
\end{figure}

We first explore whether the surface brightness distribution of  the host galaxy of \so\   can be well fitted by a  \dev\ surface brightness profile, typical of elliptical galaxies hosting gamma-ray blazar nuclei \citep{kotilainen98, falomo00}. The best fitted model parameters are listed in Table~\ref{tab:galfit}. Nonetheless, inspection of the image and luminosity profile residuals (see left panels of Figure~\ref{fig:j}) show that the host galaxy of \so\ is not adequately fitted by a single S\'ersic component.  

\begin{figure}[t]
\includegraphics[width=\textwidth]{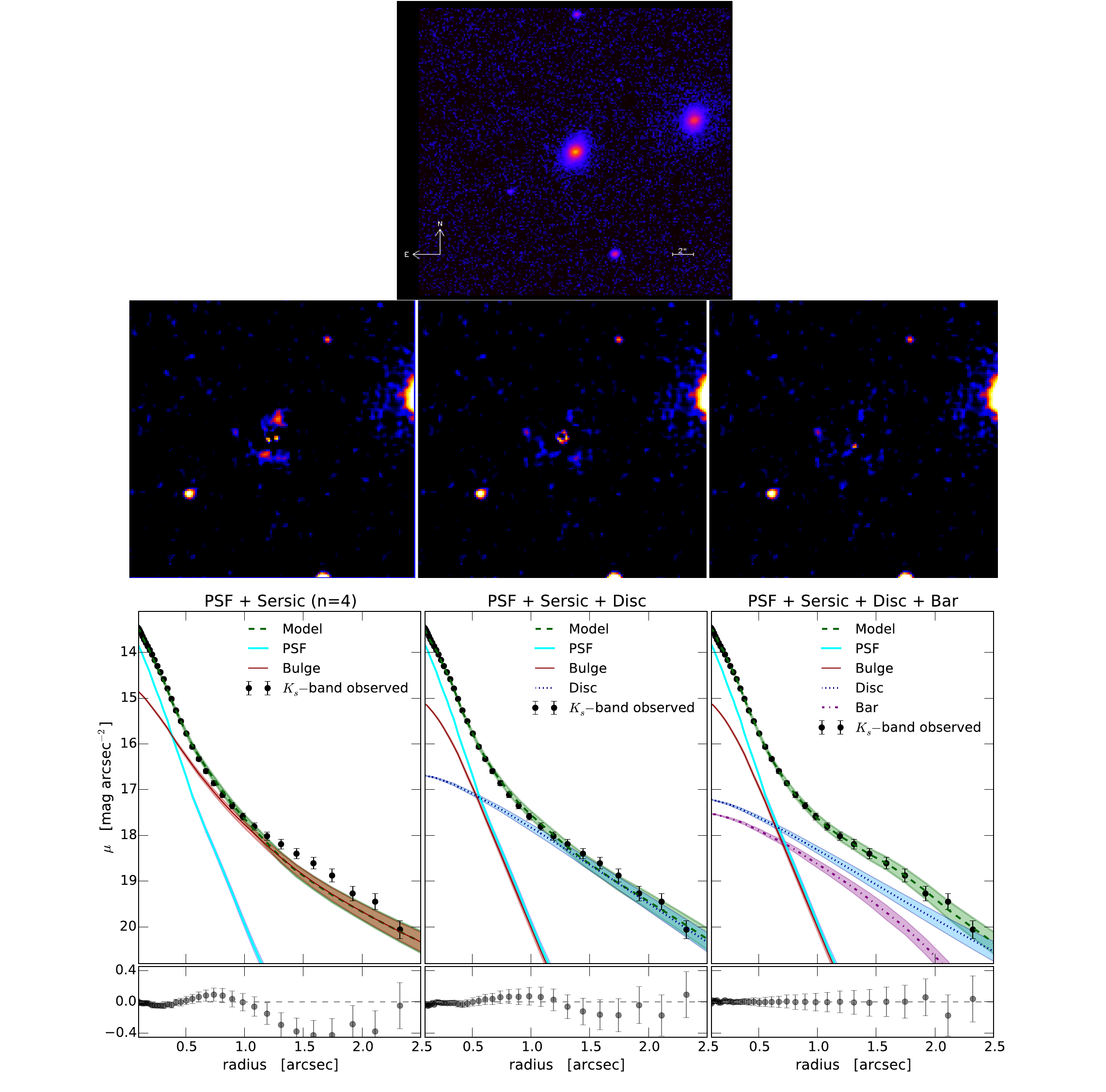}
\caption{The layout is the same as in Figure \ref{fig:j} but for $K_s-$band.}\label{fig:k}
\end{figure}

Thus, we added a component to the host galaxy model representing an exponential disc, the latter is motivated by the high incidence of radio-quiet \sy\ in late-type galaxies \citep{crenshaw03,deo06,ohta07}.  Middle panels of Figures~\ref{fig:j} and \ref{fig:k} show that the observed surface brightness distribution of \so, in each filter, can be better represented  with a model comprising  a nuclear unresolved source, a bulge and an exponential disc.  Based on the $\chi^{2}$ values listed on Table~\ref{tab:galfit}  and from an inspection of the residuals in Figures~\ref{fig:j}  and \ref{fig:k}, a host galaxy model including a bulge and an exponential disc seems to be more successful to reproduce the observed surface brightness of \so. However, the residual images in both bands (middle panels in Figures~\ref{fig:j}  and \ref{fig:k}) reveal the  presence of two aligned bright spots, consistent with the residual structure obtained with a \dev\ model (left panels).  Symmetric aligned bright knots are usually found at the ends of stellar bars, the so called ansae or handles of the bar \citep{laurikainen07}.  Stellar bars have not been detected in AGN launching prominent relativistic jets, however, bars appear rather frequent in radio-quiet \sy\ \citep{deo06,ohta07}.

\begin{figure}[t]
\includegraphics[width=\textwidth]{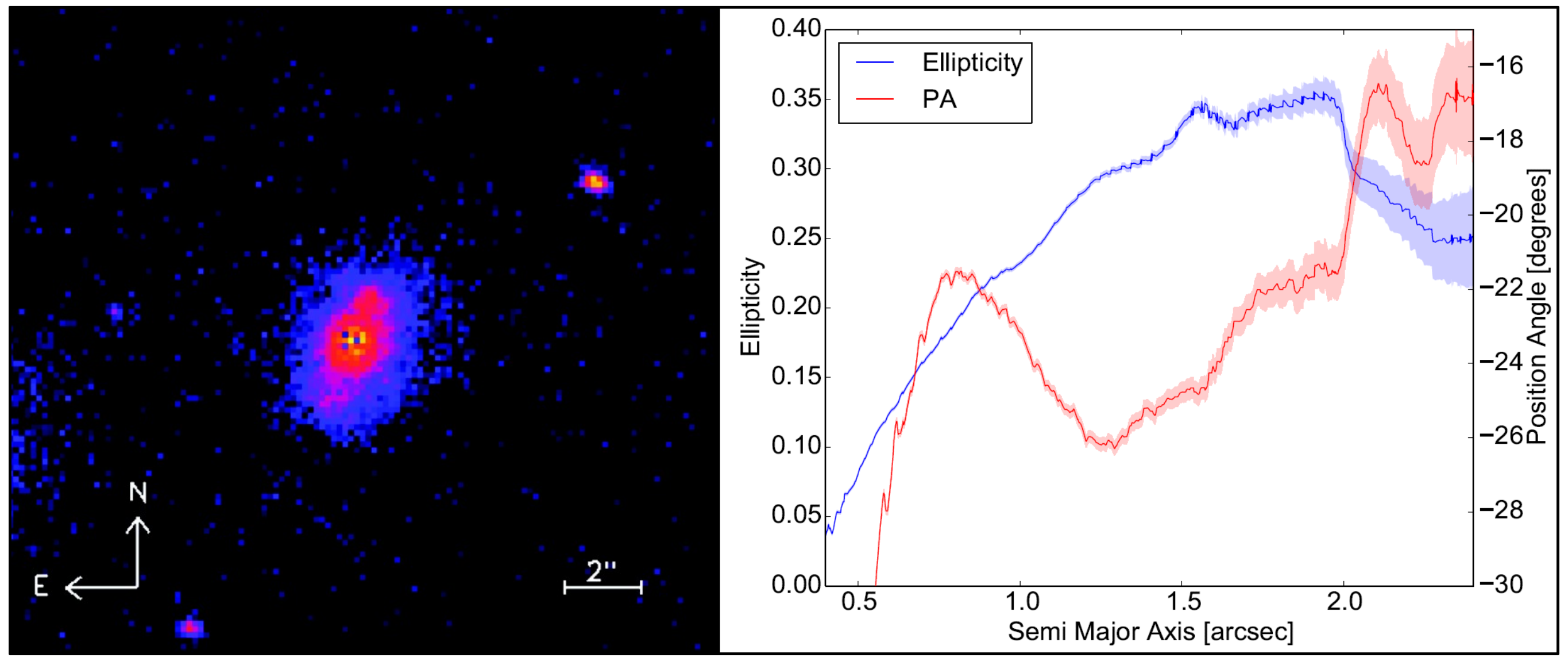}
\caption{\textit{Left panel:} Residual image obtained by subtracting the best nuclear and bulge components from the observed $J-$band images of \so. Based on our analyses, the central elongated structure has been identified as a stellar bar. \textit{Right panel:} Ellipticity and P.A. radial profiles. The ellipticity profiles shows a sharp drop around 2 arc sec matching a change in P.A. Moreover, the maximum value of ellipticity is larger than 0.25, fulfill common criteria for stellar bars identification. }\label{fig:bar}
\end{figure}

We subtract the best-fitted PSF and \ser\ models from the observed image  to search for a bar component in the host galaxy of \so. The resultant image is shown in the left panel of  Figure~\ref{fig:bar}  revealing a disc component with an elongated brightness enhancement  along the inner region. Since the S/N is lower than in the $J$-band, the $K_{s}$-band subtraction is not shown. Hence, it is tempting to visually identify such a structure with a bar.  Nevertheless, to ascertain the presence of a stellar bar in the host galaxy of \so, we have used the IRAF task \texttt{ellipse} to fit ellipses to the isophotes of the observed galaxy surface brightness,  allowing us  to extract  ellipticity and position angle (P.A.) radial profiles from our both images. The ellipticity profiles of barred galaxies should increase monotonically and  decline abruptly at the transition from the bar to the disk,  at the same location a sharp change in PA should be seen \citep{cisternas15}.  Moreover, the maximum value of ellipticity should be larger than 0.2 \citep{menendez07}.  This expected behaviour can be seen in the ellipticity profiles of \so\ shown in the right panel of Figure~\ref{fig:bar}.

Since our analysis  suggests the presence of a stellar bar,  the decomposition of the 2D \so\ host galaxy light distribution should take into account all the structural components. Therefore a four-component model consisting of a PSF  (to model the unresolved nuclear AGN emission), a bulge, an exponential disc and a bar has been fitted to  our images of \so. As is customary in studies of barred galaxies, we model the bar with a \ser\ component with low $n < 1$ \citep{laurikainen05}, hereafter we refer to this component simply as the bar. The best-fit parameters for this model are also listed in Table~\ref{tab:galfit} and its residual images and radial profiles are shown in the right panels of Figures~\ref{fig:j}  and \ref{fig:k}.  
We have explored several combinations of parameters in the later model and noticed that including a bulge component with  $n > 2$ was always resulting in poor fits showing residuals dependence with radial distance and clearly underperforming ($\chi^{2}_{J} \geq 1.470$ and  $\chi^{2}_{K_s} \geq 1.454$)  when compared to our best-fit models ($\chi^{2}_{J} = 1.405$ and  $\chi^{2}_{K_s} = 1.398$). Similarly, we tried to fit the galaxy without the bulge (due to its small size; see Table \ref{tab:galfit}). However, a considerable fraction of light (B/T$\approx0.25$) comes from the bulge, and removing it results in physically meaningless morphologies for the other components of the model (e.g. S\'ersic index for the bar $n>10$).\smallskip

  %%%%%%%%%%%%%%%%
%%%%%%%%%%%%%%%%
 %%%%%%%%%%%%%%%%

 %%%%%%%%%%%%%%%%
%%%%%%%%%%%%%%%%
 %%%%%%%%%%%%%%%%
 
  %%%%%%%%%%%%%%%%
%%%%%%%%%%%%%%%%
 %%%%%%%%%%%%%%%%
 
   %%%%%%%%%%%%%%%%
%%%%%%%%%%%%%%%%
 %%%%%%%%%%%%%%%%

\section{Discussion}

The best-fit model for the host galaxy structure of \so\ reveals it as a barred late--type disc galaxy. Given its stellar mass, estimated from its $K-$band luminosity \citep[$\sim7\times10^{11}M_\odot,$ adopting the mass--to--light ratio found by][]{mcgaugh14}, the bar feature is not as expected as it would be for a late--type disc galaxy. \cite{cameron10} find that for galaxies with stellar masses ($M_\star>10^{11}M_\odot$), the fraction of bars in early-type discs is significantly lower ($f_{bar}=0.19\pm0.03$), at all redshifts, than that in late-type discs ($f_{bar}=0.42\pm0.05$).

%\textbf{The host galaxy of \so\, also comprises of a bulge with a remarkably low \ser\ index ($n \sim 1.2$) and a bulge--to--total ratio  \citep[B/T$=0.56\pm0.02$ and B/T$=0.60\pm0.02$ for J-- and Ks--band, respectively without taking into account the AGN contribution and adding the bar luminosity to the bulge, according to][]{Gadotti_2009}. These values are well within the characteristics of pseudobulges \citep{kormendy_2004}.

The host galaxy of \so\, also comprises of a bulge with a remarkably low \ser\ index ($n \sim 1.2$) and a bulge--to--total ratio (B/T$=0.39\pm0.02$ and B/T$=0.44\pm0.03$ for J-- and Ks--band, respectively --without taking into account the AGN contribution--). These values are well within the characteristics of pseudobulges \citep{kormendy04}. We can further explore whether \so\ can be classified as having a pseudobulge by comparing the results presented in \citet{fisher08} with the values in our analysis (see Table \ref{tab:galfit}) that do not depend on magnitude \citep[since they seem to be independent of wavelength,][]{graham01,macarthur03,fisher08}. In the $R_{eff}/R_s$ distribution for the bulges of the galaxies in their sample, they find that classical bulges show a $\langle R_{eff}/R_s \rangle=0.45\pm0.28$, whereas pseudobulges show a $\langle R_{eff}/R_s \rangle=0.21\pm0.10$, in accordance with the (pseudo) bulge in \so\ ($\langle R_{eff}/R_s \rangle=0.20\pm0.11$). In their figures 7 and 11, they show the planes $n-R_{eff}$ and $n-\epsilon_{bulge}/\epsilon_{disc}$, respectively, where it is evident that the (pseudo) bulge in \so\ lies within the psedobulge subsample. All the latter suggests that the bulge in \so\ should not be classified as classical but as pseudo. Unlike classical bulges, commonly found in blazars and radio galaxies (grown trough mergers), pseudobulges are built slowly by the internal evolution of the galaxy, likely driven by the bar. For a full and detailed description on pseudobulges see \cite{kormendy04}.\smallskip

A rapidly spinning black hole by-product of a BH--BH merger is thought to play a fundamental role in triggering of radio loud activity in AGN  and the launching of relativistic jets from supermassive black holes \citep{blandford77}. However, the firm detection of a pseudobulge and a stellar bar in the host galaxy of \so\ argue for a scenario where an instability driven bulge has concomitantly grown with its black hole via secular processes -- albeit some minor galaxy mergers could have occurred. Only a small number of powerful radio galaxies have been found in disc--like host systems \citep[e.g.][]{ledlow98,hota11,morganti11,bagchi14,kaviraj15,mao15,singh15}, and among them, PKS 2004-447 is the only gamma-ray emitting radio galaxy in which a disky pseudobulge and a stellar bar have been found.

Since there is solid observational evidence pointing to the presence of fully developed relativistic jets emanating from \so\ \citep{orienti15,schulz16},  it arises that the same spinning black hole conditions that enable the production of prominent relativistic jets can be met,  even if  non-major merger related black hole fueling mechanisms  have ignited the AGN.  Our results allow simulations and subgrid models to explore the energetic feedback of a radio jet to an interstellar medium different from an elliptical galaxy \citep{wagner11} and provide tests on black hole spin up mechanisms that are not strictly associated to major BH--BH mergers.

A recent spectropolarimetric observation of \so\ has revealed a broad and double-peaked H$\alpha$ profile in polarised light \citep{baldi16}, this peculiar profile was interpreted as the signature of an obscured disc-like broad-line region. Estimating the black hole mass of \so\  from a single-epoch polarised spectrum yields a black hole mass  ($\sim 6 \times 10^{8}$M$_{\odot}$)  well within the range  of radio-loud AGN \citep{dunlop03}.  However, double-peaked Balmer emission line profiles are highly variable \citep{lewis10}. Thus,  bulge stellar kinematics measurements are preferred over emission line widths to estimate black hole masses in double-peaked AGN \citep{lewis06}.

Moreover, non-merger driven black hole fueling mechanisms  are known to be  inefficient, hence, setting an upper limit to the achievable mass of the black hole  \citep{hopkins09}, well within the common range of black hole masses estimated for  NLSy1 galaxies ($\sim \times 10^{7}$ M$_{\odot}$). Therefore, a solid estimation of the black hole mass of \so\ based on its polarised broad emission  line profile would require further investigation on whether motions of ionised gas are dominated by gravity or  if outflowing winds \citep{longinotti15}, could contribute to shaping the double-peaked profile. We, however, notice that black holes with masses surpassing 10$^{8}$ M$_{\odot}$ have  also been found in pseudobulges  \citep{kormendy13}. Moreover, using the K--band bulge luminosity ($L_{bulge}$) from our analysis and the $M_{BH}-L_{bulge}$ relation of pseudo-bulges from \citet{ho14}, we estimate the black hole mass in \so\ to be $\sim 9 \times 10^{7}$M$_{\odot}$. Nevertheless, our imaging study provides observational evidence for a significant difference between  \so\ and  the overall population of gamma-ray emitting  AGN in terms of evolutionary stages of their  host galaxies.

We have found that \so\ is powered by a suspected ongoing growing black hole hosted by a galaxy that is actively evolving, albeit slowly via secular processes. Although there is ample observational  evidence linking the radio-loud AGN phenomenon to mergers \citep{ramosalmeida12,chiaberge15}, our results suggest that the growth of a black hole via hierarchical clustering (major mergers) might not be strictly required to  launch fully developed relativistic jets. Thus, there is an urgent need for more observations to characterise the evolutionary state and nuclear properties of the six NLSy1s detected by Fermi so far. To investigate this further, we have recently carried out a NIR host galaxy imaging of all the gamma-ray NLSy1s known to date and a large sample of radio-loud NLSy1s, to allow comparison between populations.

\section{Summary}

We  have used the NIR ISAAC camera on the VLT to image for the first time the host galaxy of \so\ -- one of the six $\gamma$-ray \nls\ galaxies detected by the \emph{Fermi Gamma-ray Space Telescope}.  The surface brightness distribution of the host galaxy  of \so\ has been decomposed with GALFIT, allowing us to characterise its structural galaxy components.  Our main results  are listed below:

\begin{itemize}

\item We find that the surface brightness distribution of  host galaxy of \so\ deviates significantly from a \dev\ profile, hence, does not share galaxy morphological features with the overall population of gamma-ray bright AGN (FSRQs, BLLacs and radio galaxies).

\item We perform a 2D bulge-disk decomposition on the images of \so, yielding to better results. However, the presence of two aligned brightness enhancements in the residual images (see Figures~\ref{fig:j} and \ref{fig:k})  were further identified as the ansae or handles of a stellar bar. 

\item As a consequence, a 2D bulge-disk-bar decomposition turned out to be the best representation of the host galaxy of \so. Most of the observational properties of the best fitted bulge (e.g.  $n < 2$) are consistent with those reported in literature for pseudobulges.  

\so\ arises therefore as the first gamma--ray emitting radio--loud AGN where a fully developed relativistic jet -- able to accelerate particles up to gamma-ray energies --  is launched from a pseudo bulge grown via secular processes, and not through mergers. 

\end{itemize}

%\bibliography{ms_10may16}

\end{document}